\documentclass[sigconf,natbib=false]{acmart}
\AtBeginDocument{%
  }

\RequirePackage[
  datamodel=acmdatamodel,
  style=acmnumeric,
  ]{biblatex}

\usepackage{graphicx}
\usepackage{caption}
\usepackage{multirow}
\usepackage{multicol}
\usepackage{array}
\usepackage{siunitx}
\usepackage{pgfplots}
\pgfplotsset{compat=1.18}
\usepackage{pgfplotstable}
\usepackage{xcolor}
\usepackage{subcaption} 
\usepackage{hyperref}
\usepackage[most]{tcolorbox}
\begin{document}

\title{Seed\&Steer: Guiding Large Language Models with Compilable Prefix and Branch Signals for Unit Test Generation}
% \title{Automating Unit Test Generation with Compilable Prefixes and Branching Signals Using Large Language Models}
% \title{PromptWeaver: Weaving Compilable Prefix and Branch Signals for Precise LLM Unit Test Generation}
% \title{EvoBranchPilot: Steering LLMs with Compilable Seeds and Conditional Branches for Test Generation}

\author{Shuaiyu Zhou}
\affiliation{%
  \institution{Peking University}
  \city{Beijing}
  \country{China}}
\email{monster290@stu.pku.edu.cn}

\author{Zhengran Zeng}
\affiliation{%
  \institution{Peking University}
  \city{Beijing}
  \country{China}}
\email{zhengranzeng@stu.pku.edu.cn}

\author{Xiaoling Zhou}
\affiliation{%
  \institution{Peking University}
  \city{Beijing}
  \country{China}}
\email{xiaolingzhou@stu.pku.edu.cn}

\author{Rui Xie}
\affiliation{%
  \institution{Peking University}
  \city{Beijing}
  \country{China}}
\email{ruixie@pku.edu.cn}

\author{Shikun Zhang}
\affiliation{%
  \institution{Peking University}
  \city{Beijing}
  \country{China}}
\email{zhangsk@pku.edu.cn}

\author{Wei ye}
\affiliation{%
  \institution{Peking University}
  \city{Beijing}
  \country{China}}
\email{wye@pku.edu.cn}

\begin{abstract}
Unit tests play a vital role in the software development lifecycle. Recent advances in Large Language Model (LLM)-based approaches have significantly improved automated test generation, garnering attention from both academia and industry. We revisit LLM-based unit test generation from a novel perspective by decoupling prefix generation and assertion generation. To characterize their respective challenges, we define Initialization Complexity and adopt Cyclomatic Complexity to measure the difficulty of prefix and assertion generation, revealing that the former primarily affects compilation success, while the latter influences test coverage.

To address these challenges, we propose \textit{Seed\&Steer}, a two-step approach that combines traditional unit testing techniques with the capabilities of large language models. \textit{Seed\&Steer} leverages conventional unit testing tools (e.g., \texttt{EvoSuite}) to generate method invocations with high compilation success rates, which serve as seeds to guide LLMs in constructing effective test contexts. It then introduces branching cues to help LLMs explore diverse execution paths (e.g., normal, boundary, and exception cases) and generate assertions with high coverage.
We evaluate \textit{Seed\&Steer} on five real-world Java projects against state-of-the-art baselines. Results show that \textit{Seed\&Steer} improves the compilation pass rate by approximately 7\%, successfully compiling 792 and 887 previously failing cases on two LLMs. It also achieves up to ~73\% branch and line coverage across focal methods of varying complexity, with coverage improvements ranging from 1.09× to 1.26×. Our code, dataset, and experimental scripts will be publicly released to support future research and reproducibility.
\end{abstract}


% \received{18 Friday 2025}
% \received[revised]{XX XX 2025}
% \received[accepted]{XX XX 2025}

\maketitle

\section{Introduction}
Unit testing is a key part of modern software development, used to check individual methods or functions before system integration \cite{olan2003unit, runeson2006survey, zhu1997software}. However, manually writing and maintaining these tests is time-consuming and error-prone \cite{daka2014survey,klammer2015writing}, especially as codebases evolve. To reduce this workload, researchers have developed automated test generation techniques—including search-based \cite{blasi2022call, delgado2022interevo, harman2009theoretical, baresi2010testful, derakhshanfar2022basic, fraser2010mutation, harman2001search}, constraint-based \cite{csallner2008dysy, ernst2007daikon, xiao2013characteristic, ma2015grt, sakti2014instance}, and random-based \cite{zeller2019fuzzing,pacheco2007feedback,andrews2011genetic} approaches—which mainly aim to maximize code coverage.

However, these traditional methods often create tests that are hard to read, understand, or reuse, limiting their practical adoption \cite{almasi2017industrial, gargari2021sbst}. Language model-based approaches have emerged as a promising alternative. Using pretrained Transformer architectures \cite{vaswani2017attention}, tools like A3test \cite{alagarsamy2024a3test} and AthenaTest \cite{tufano2020unit} can learn from human-written tests to generate cases that are structurally correct and logical.
Furthermore, large language models like ChatGPT \cite{achiam2023gpt} show a strong ability to generalize for code understanding and generation, approaching human-level comprehension \cite{feng2024improving, nam2024using, saito2024unsupervised}. Similarly, LLM-based tools such as TestGen-LLM \cite{alshahwan2024automated}, \cite{bhatia2024unit}, ChatUniTest \cite{xie2023chatunitest}, and ChatTESTER \cite{yuan2023no} can produce easy-to-understand tests with good accuracy and coverage. These LLM-based approaches have significantly advanced unit test generation, attracting widespread attention in both academia and industry.

\begin{figure*}[t]
  \centering
  \begin{subfigure}[b]{0.48\textwidth}
    \centering
    \includegraphics[width=\textwidth]{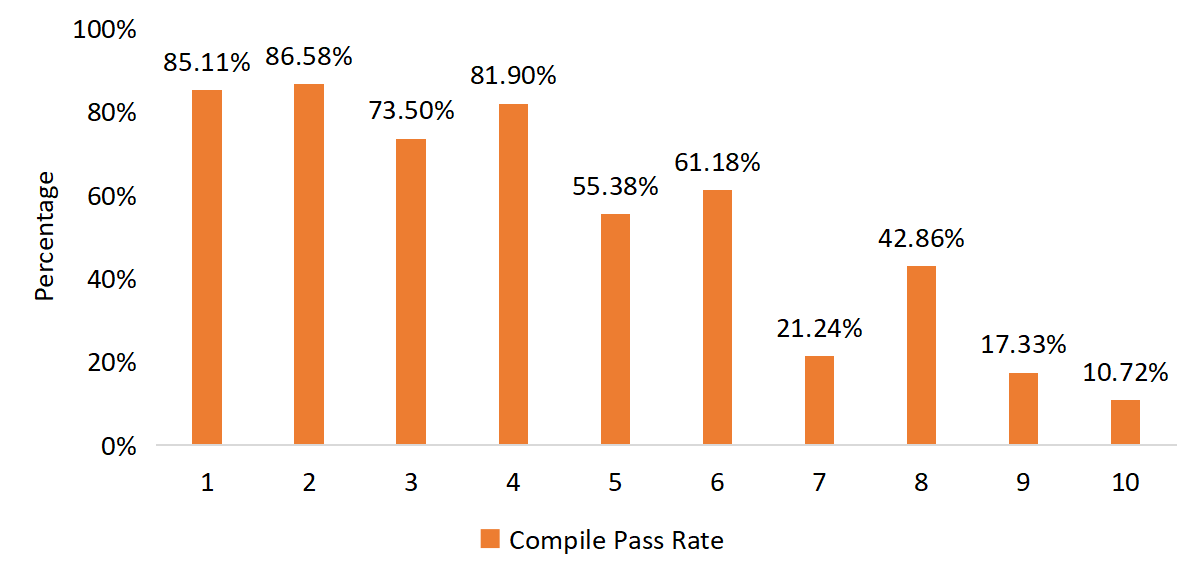}
    \caption{Initialization Complexity}
    \label{fig:init_complexity}
  \end{subfigure}
  \hfill
  \begin{subfigure}[b]{0.48\textwidth}
    \centering
    \includegraphics[width=\textwidth]{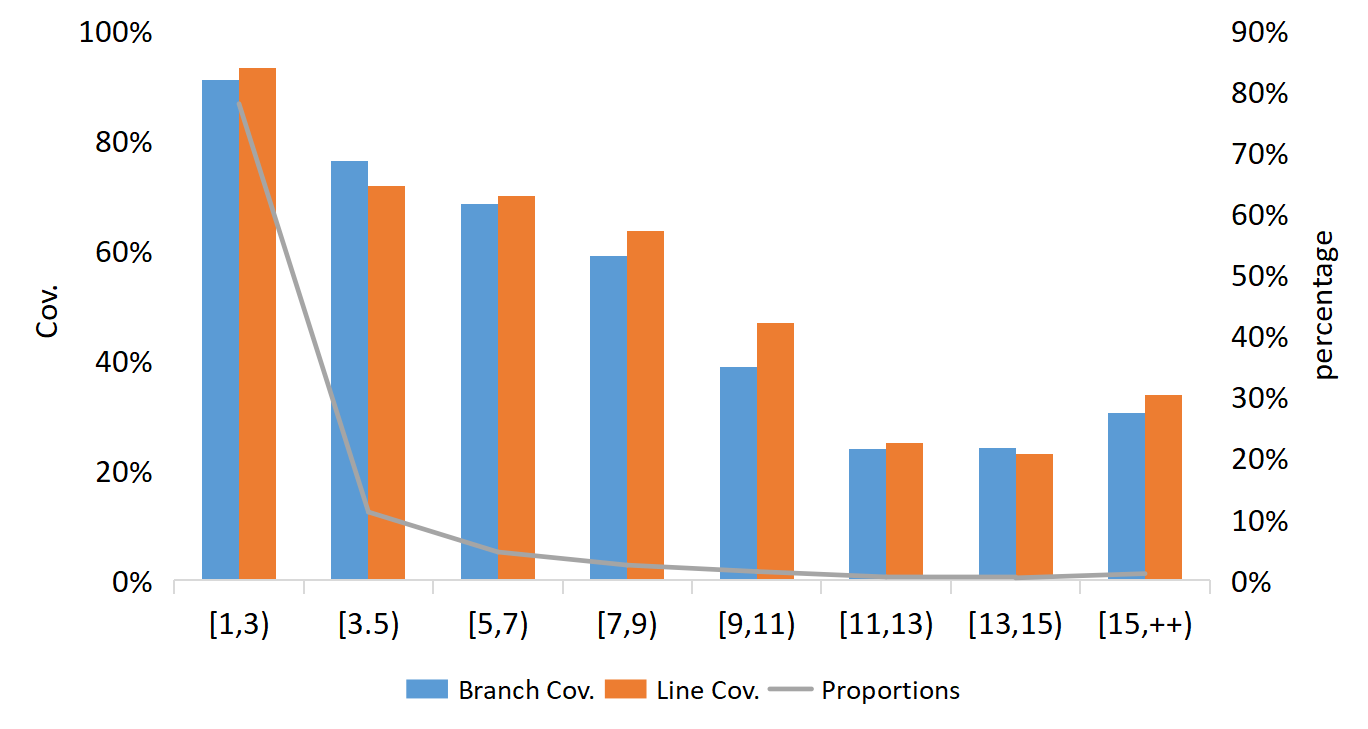}
    \caption{Cyclomatic Complexity}
    \label{fig:cc_complexity}
  \end{subfigure}
  \caption{(a)With higher initialization complexity, LLMs struggle to invoke the focal method correctly. Correct invocation examples may help LLMs learn to generate compilable prefixes. (b)As cyclomatic complexity grows, simple instructions fail to convey code context. Guiding LLMs with more branch information may lead to deeper coverage.}
  \label{fig:complexity_distribution}
\end{figure*}

Unit test generation consists of two main phases: test prefix generation and assertion generation. The test prefix sets up the testing environment by building the test class structure, initializing variables, and configuring object states to ensure the target method runs correctly. Assertion generation verifies the method’s correctness by creating conditions that compare actual results with expected outcomes. In this work, we examine LLM-based unit test generation from a novel, finer-grained perspective by decomposing the process into these two distinct stages. We define Initialization Complexity and adopt Cyclomatic Complexity to assess the difficulty of prefix and assertion generation. Through preliminary studies (detailed in Section~\ref{Preliminary Study}), we reveal that test prefix generation and assertion generation have distinct impacts on the compilability and coverage of the final test cases. Specifically, we present two key findings:

\textbf{(1) The lack of proper method invocation in the prefix generation stage is a major contributor to compilation failures}. Many real-world methods require complex initialization contexts, such as nested object creation, precondition configuration, or specific method invocation sequences. Due to limited understanding of domain-specific method usage, LLMs often generate invalid test prefixes, leading to uncompilable code. As discussed in Section~\ref{initialization}, we found that the higher the initialization complexity (see formula~\ref{eq:init_complexity}) of a method, the more difficult it is for LLMs to generate compilable tests, with success rates dropping to as low as 10.72\% in the most complex cases.

\textbf{(2) Complex code logic causes generated assertions to be superficial, leading to poor branch and line coverage}. Existing prompting strategies often rely on generic instructions such as "write a test case for this method and follow these rules: ...", which offer only coarse-grained guidance. This makes it difficult for LLMs to reason about fine-grained branching conditions or boundary cases, especially in methods with non-trivial control flow. As discussed in Section~\ref{Cyclomatic}, we found that for methods with complex code logic, LLMs struggle to fully understand the context, resulting in limited coverage of deeper parts of the method, sometimes covering only around 20\% code in the worst cases.

In this paper, we present \textit{Seed\&Steer}, a novel two-step approach that combines traditional unit testing practices with the capabilities of large language models. In the test prefix generation step, which we named the Seed process, we introduce \texttt{EvoSuite}, which analyzes almost every method in a test class, including local methods, based on bytecode. We extract the method invocation cases from them to help LLMs generate unit tests, which compensates for LLMs' limited understanding of class constructors, object lifecycles, and dependency constraints by generating test cases that successfully invoke the focal method. In the assertion generation step (Steering process), we first extract conditional branching statements in the focal method to assist the LLM in generating branching intents, which help the LLM focus on different execution paths (e.g., normal, boundary, and exception cases), and finally generate assertions in conjunction with function intents. We conduct extensive experiments on five real-world Java projects and demonstrate that \textit{Seed\&Steer} consistently outperforms existing baselines in terms of compilation success, test execution, as well as branch and line coverage.

In summary, our work makes the following key contributions:
\begin{itemize}
\item We revisit LLM-based unit test generation from a novel perspective by separating prefix generation and assertion generation, revealing that the former primarily affects compilation success, while the latter influences test coverage.

\item We propose \textit{Seed\&Steer}, a simple yet effective method that combines traditional tools with LLMs to construct compilable test prefixes and generate branch-guided assertions, achieving promising results in compilation success, test execution, and code coverage.

\item  Our findings categorize the limitations of LLM-based automated unit test generation into two concrete complexity challenges, which offers a unique perspective for future research.
\end{itemize}

We will release all code, datasets, and experiment scripts \footnote{\url{https://github.com/monster29000/SeedandSteer}} to support reproducibility and foster future research in LLM-based unit test generation.

\begin{figure*}[t]
  \centering
  \includegraphics[width=\textwidth]{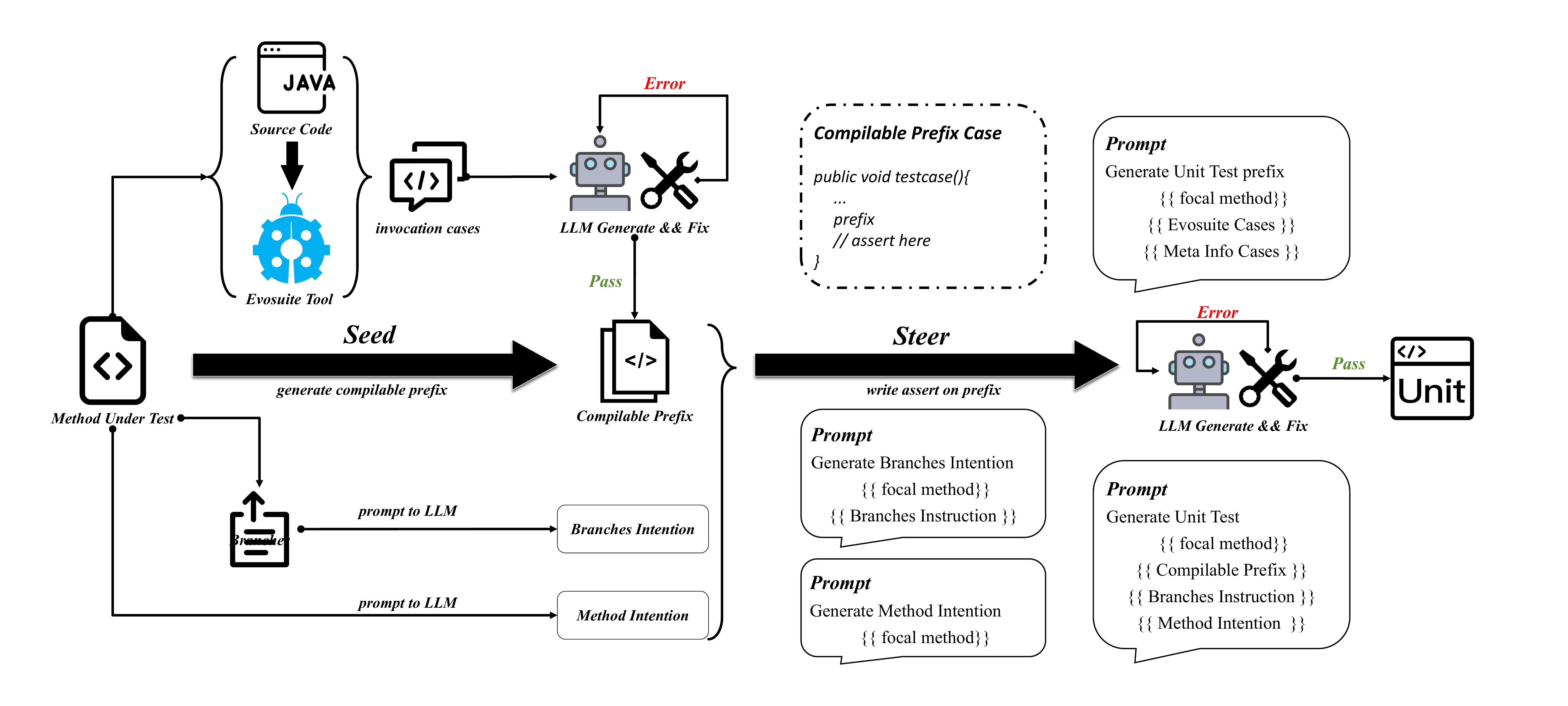}
  \caption{Overview of the Seed\&Steer workflow.}
  \Description{A diagram showing the seed prefix and branch extraction feeding into an LLM prompt, followed by test generation, compilation check, and correction.}
  \label{fig:overview}
\end{figure*}

\section{Preliminary Study}
\label{Preliminary Study}
We found that there are always limitations when generating unit tests for complex focal methods. To better illustrate the challenges we found, we quantified the impact of \textbf{Initialization Complexity} and \textbf{Cyclomatic Complexity} on the existing method \textit{ChatTester}~\cite{yuan2023no}.

\subsection{Initialization Complexity}
\label{initialization}
To quantify the impact of initialization logic on test generation effectiveness, we define four static features: the number of variable declarations $V$, the number of object creations $O$, the number of method calls before invoking the focal method $M$, and parameter number $P$. Based on these, We define the overall \textbf{Initialization Complexity Score} as a weighted sum of four normalized static features:
\begin{equation}
\text{Initialization Complexity} = w_1 \cdot \widehat{V} + w_2 \cdot \widehat{O} + w_3 \cdot \widehat{M} + w_4 \cdot \widehat{P}
\label{eq:init_complexity}
\end{equation}
Here, $\widehat{x}$ denotes the min-max normalized value of feature $x$, and $w_1$, $w_2$, $w_3$, $w_4$ represent the corresponding weights. Through correlation analysis, we observed that the number of method calls before invoking the focal method $M$ and the number of parameters $P$ are more strongly negatively correlated with compilation success. Therefore, we assign higher weights to these two factors in the final complexity score, resulting in a weight configuration of $[0.1, 0.1, 0.4, 0.4]$.

As shown in Figure~\ref{fig:init_complexity}, we scale the Initialization Complexity Score into bins (0–1, 1–2, ..., 9–10) and report the compilation pass rate within each bin. The trend is clear: as the initialization complexity increases, the proportion of tests that are successfully compiles drops significantly. For instance, tests with complexity scores between 1 and 2 achieve a pass rate of 86.58\%, while those in the 9–10 range drop to only 10.72\%. This observation highlights that initialization complexity is a major obstacle in generating compilable test code. This suggests that the complex initialization  critical obstacle to compilation success. 

We find that complex initialization is a key limiting factor in automated test generation. As initialization becomes more intricate, LLMs often fail to correctly invoke the focal method, leading to compilation failures.

\subsection{Cyclomatic Complexity}
\label{Cyclomatic}
Beyond initialization logic, we also investigate the impact of focal method structural complexity on test generation. We compute the Cyclomatic Complexity Number (CCN) of each focal method using the static analysis tool \texttt{Lizard}, and correlate it with coverage metrics.

As shown in Figure~\ref{fig:cc_complexity}, we observe that approximately 85\% of methods have a CCN between 1 and 3, typically representing simple constructors or utility functions. These low-complexity methods are generally well-handled by state-of-the-art large language models. In contrast, methods with CCN values ranging from 4 to 10 often containing intricate control flow and business logic-exhibit a sharp decline in test coverage, with both branch and line coverage typically falling below 50\%.

We find that method complexity is a limiting factor in automated test generation. As code logic becomes more complex, large language models often fail to fully understand the code context, resulting in lower code coverage.

\section{Methodology}
This section presents a detailed introduction to the proposed method, \textbf{\textit{Seed\&Steer}}. The goal of \textit{Seed\&Steer} is to enhance the effectiveness of LLMs in unit test generation, whose core idea is to divide the test generation process into two key stages. In the first stage, we collect representative invocation cases of the focal method and construct high-quality, compilable test prefixes that serve as a stable and semantically meaningful foundation for subsequent test generation. In the second stage, we extract conditional branch signals from the focal method to derive branch intents, which are then integrated with functional intents to formulate more precise and context-aware prompts. These prompts guide the model in generating assertions with improved semantic alignment and higher code coverage.

In the following subsections, we initially elaborate on the extraction of meta information from static source code in Section~\ref{sec1} and then discuss the construction of compilable test prefixes in Section~\ref{sec2}. Furthermore, Section~\ref{sec3} introduces our strategy for leveraging branching intents to guide the generation of high-quality assertions. Finally, Section~\ref{sec4} details the repair mechanism and the iterative generation strategy designed to enhance test effectiveness.

\subsection{Code Preprocessing}
\label{sec1}
Before initiating the test generation process, we conduct a static parse over the source code to extract structured meta information that will be utilized in subsequent stages, such as prefix construction and branch guidance.

Specifically, we traverse each Java file in the target project and parse it into an Abstract Syntax Tree using the \texttt{javalang} and \texttt{tree-sitter} libraries in Python. This parsing enables us to systematically extract class-level information, including class name, type (e.g., interface, abstract, or concrete), inheritance relations, source code content, field declarations, and all method definitions.

We formalize a class as a structured object $\mathcal{C}$, defined as:
\[
\mathcal{C} = \{ \text{name}, \text{type}, \text{superclass}, \text{content}, \text{fields}, \mathcal{M} \},
\]
where $\mathcal{M}$ denotes the set of methods declared in the class. We further partition $\mathcal{M}$ into two subsets:

\begin{itemize}
  \item $\mathcal{M}_\text{focal}$: Public methods that are considered as targets.
  \item $\mathcal{M}_\text{other}$: Non-focal methods such as private or utility.
\end{itemize}

A focal method is defined as a public method in a public, non-abstract class. Accordingly, we define the focal method set $\mathcal{F}$ as follows:
\[
\mathcal{F} = \{ f_i = (\text{name}, \text{params}, \text{invocation}, \text{signature}, \text{content}) \mid f_i \in \mathcal{M}_\text{focal} \}.
\]

Each focal method $f_i$ contains its name, parameter list, invocation information (i.e., which methods it calls), full signature, and the corresponding method code. The "invocation" in the formula is used to filter the correct call cases of the function, the "signature" is utilized to filter the focal method when calculating the coverage, and the content is employed to extract the conditional branching statements. This structured metadata is essential for building meaningful prompts and constructing compilable test prefixes in the downstream test generation pipeline.

This systematic preprocessing process ensures that the model is provided with precise and complete contextual information, which serves as the foundation for the subsequent \textbf{steps}:  \textit{Seed} and \textit{Steer}.

\subsection{Seed: Compilable Prefix}
\label{sec2}
To generate high-quality test prefixes for a given focal method $\mathcal{M}_\text{focal}$, we design a seed generation strategy that combines static parse with example extraction. This strategy leverages both automatically generated tests from EvoSuite and manually discovered calling patterns from the source class $\mathcal{C}$ to ensure realistic and executable prefix construction.

\textit{Generating EvoSuite test set} $\mathcal{T}_{ES}$.
We begin by applying \texttt{EvoSuite}, a bytecode-based unit test generator, to class $\mathcal{C}$ that contains the focal method. \texttt{EvoSuite} generates a compilable and structurally valid test class, where each generated test class is processed into a structured representation:
\[
\mathcal{T}_{ES} = \{ \text{name}, \text{type}, \text{content}, \text{fields}, \mathcal{M}_{tc}\},
\]
where $\mathcal{M}_{tc}$ denotes the set of test methods, each of which may include one or more method invocation traces. All test code is preprocessed using the same procedure described in Section~\ref{sec1}.

Subsequently, to enhance the executability and realism of the prefixes, we adopt a dual-path strategy:

\textit{Path 1: Source-level exemplar mining.}
We traverse all methods in the source class, denoted by $\mathcal{M} = \mathcal{M}_\text{focal} \cup \mathcal{M}_\text{other}$. For each method $m \in \mathcal{M}$, if the method's invocation list contains a call to $\mathcal{M}_\text{focal}$, we extract the object instantiation and argument preparation statements as a candidate prefix template.

To further improve diversity, we additionally examine the EvoSuite-generated test methods $\mathcal{M}_{tc}$, and if any method invokes $\mathcal{M}_\text{focal}$, we extract at most three such test cases.

\textit{Path 2: EvoSuite-only mining when source lacks calls.}
If no method in $\mathcal{M}$ directly invokes the focal method, we fallback to the EvoSuite test cases. We iterate over all $m \in \mathcal{M}_{tc}$ in $\mathcal{T}_{ES}$ and extract up to five examples where the focal method is invoked.

\textit{Constructing LLM prompts and generation.} We then construct a prompt that contains the extracted positive examples and explicitly instruct the LLM to generate only the test prefix that correctly invokes $\mathcal{M}_\text{focal}$, without assertions. Upon generation, we remove all assertion statements and inject comments (e.g., \texttt{// TODO: assert here}) to mark where an assertion would be added.

\textit{Iterative refinement via compilation feedback.} The generated code is compiled using a local Java compiler. If the compilation fails, we capture the error messages and construct new prompts containing the diagnostics to guide the LLM in fixing the code. This process is repeated for at most five iterations to control resource usage and token budget.

\subsection{Steer: Branch Intentions}
\label{sec3}
After obtaining a compilable test prefix $\mathcal{P}_{\text{pass}}$, the next step \textbf{Steer} is to guide the LLM to generate meaningful and effective assertions. This phase focuses on maximizing code coverage by steering the model with structured semantic and control-flow information.

We first perform a static parse on the focal method $\mathcal{M}_{\text{focal}}$ to identify key branching structures. These include:

\begin{itemize}
  \item Conditional branches: \texttt{if}, \texttt{else if}, \texttt{switch}.
  \item Loop constructs: \texttt{for}, \texttt{while}, \texttt{do-while}.
  \item Exception handling: \texttt{try-catch}, \texttt{throw}.
  \item Input-dependent statements.
\end{itemize}

The extracted set of such branch points is denoted as $\mathcal{B}_{\text{cond}}$. For each $b_i \in \mathcal{B}_{\text{cond}}$, we prompt the LLM to infer and articulate its triggering conditions in natural language. For instance, a branch like \texttt{if (a > 0 \&\& b == null)} would yield an intention such as: \textit{"This branch is taken when the first argument is positive and the second is null."} These elements collectively contribute to the formation of the branch intention $\mathcal{I}_{\text{branch}}$.

Beyond branch-level semantics, we further summarize the overall intent of the focal method through a function-level intention $\mathcal{I}_{\text{func}}$, benefiting from the success of \textit{ChatTester}~\cite{yuan2023no}. This includes the method's purpose, input-output behavior, side effects, and corner cases.

The final prompt provided to the LLM is constructed by concatenating the following components:
\begin{itemize}
  \item The verified test prefix $\mathcal{P}_{\text{pass}}$.
  \item Branch intentions $\mathcal{I}_{\text{branch}} = \{b_i\}_{i=1}^n$.
  \item The function intention summary $\mathcal{I}_{\text{func}}$.
\end{itemize}

This structured guidance enables the LLM to generate the final test case 
$\mathcal{T}_{\text{gen}}$ with precise and semantically aligned assertions. In contrast to vague natural language instructions, our approach grounds assertion generation in concrete program semantics, thereby enhancing both coverage and correctness.

\subsection{Iterative Error Correction}
\label{sec4}

After generating a complete test case, we first apply lightweight syntactic and structural corrections, including inserting missing \texttt{import} statements, renaming classes or methods, and resolving common formatting issues. This preprocessing step ensures that the test case is syntactically valid and compilable.

Subsequently, we compile and execute the generated test case in accordance with the procedure outlined in~\cite{yuan2023no}. In the event of compilation or runtime errors, we utilize the resulting error feedback to reconstruct the prompt and re-invoke the LLM for correction. This iterative refinement process continues until the errors are resolved or a predefined maximum number of attempts $\delta$ is reached.

The errors are categorized into two types:

\begin{itemize}
  \item \textbf{Compilation Errors} ($\mathcal{E}_{\text{compile}}$): Usually caused by incorrect variable usage, method signatures, or misinterpreted seed code. We extract the compiler's diagnostic messages to construct targeted prompts for regeneration.
  \item \textbf{Runtime Errors} ($\mathcal{E}_{\text{runtime}}$): These include assertion failures or exceptions such as \texttt{NullPointerException}. When assertions fail, we treat them as candidate signals for revision. Moreover, if the error persists after $\delta$ attempts, we mark the test as partially valid and still record its coverage.
\end{itemize}

Furthermore, we formalize the correction process as a repair function:
\[
\text{Fix}(\mathcal{T}, \mathcal{E}, \delta) \rightarrow \mathcal{T}^* \text{ or } \perp
\]
where $\mathcal{T}$ is the original test, $\mathcal{E}$ denotes error feedback, and $\mathcal{T}^*$ is the corrected test. If all $\delta$ attempts fail, we return $\perp$ to indicate unrecoverable failure.

In cases where the three-stage pipeline,
\[
\textit{Seed Construction} \rightarrow \textit{Intention Steer} \rightarrow \textit{Iterative Generate\&Repair}
\]
fails to generate a valid test, we fall back to a two-stage strategy:
\[
\textit{Seed Template \& Intention Steer} \rightarrow \textit{Iterative Generate\&Repair}
\]

This feedback-driven refinement process significantly improves test success rates and reduces reliance on manual inspection.

\section{Experiments}
\subsection{Datasets}
In this section, we first introduce the construction and statistics of the evaluation dataset. Afterwards, we introduce the configuration for generating tests and baselines.

\begin{table}[htbp]
    \centering
    \caption{Projects dataset}
    \begin{tabular}{ccccc}
    \hline
    Project & Version & Focal classes & Focal methods \\
    \hline
    Gson & 2.10.1 & 52 & 378 \\
    Commons-Lang & 3.1.0 & 87 & 1728 \\
    Commons-Cli & 1.6.0 & 20 & 177 \\
    Commons-Csv & 1.10.0 & 8 & 137 \\
    JFreeChart & 1.5.4 & 499 & 5772 \\
    \hline
    Total &  & 666 & 8192 \\
    \hline
    \end{tabular}
    \label{datasets}
\end{table}

\begin{table*}[htbp]
\centering
\caption{Effectiveness comparison of Seed\&Steer and baseline approaches across four standard metrics.}
\label{tab:Effectiveness comparison}
\renewcommand{\arraystretch}{1.2}
\begin{tabular}{lcccccc}
\toprule
\textbf{Method} & \textbf{Projects} & \textbf{Focal methods} & \textbf{Compile passed Rate} & \textbf{Test Passed Rate} & \textbf{Branch Coverage} & \textbf{Line Coverage}\\
\midrule
ChatGPT-3.5 & \multirow{9}{*}{Total} & \multirow{9}{*}{8192} & 74.39\% & 49.91\% & 43.10\% & 42.58\%\\
ChatGPT-4.0 & & & 80.50\% & 59.75\% & 51.86\% & 50.88\%\\
ChatUnitTest & & & 76.52\% & 60.05\% & 48.68\% & 47.39\%\\
TestART & & & 87.05\% & \textbf{78.55\%} & 69.40\% & 68.17\%\\
ChatTester\textsubscript{\texttt{gpt-3.5-turbo}} & & & 85.88\% & 57.18\% & 50.80\% & 48.68\%\\
ChatTester\textsubscript{\texttt{gpt-4o}} & & & 87.21\% & 52.12\% & 52.78\% & 53.52\%\\
Seed\&Steer\textsubscript{\texttt{gpt-3.5-turbo}} & & & \underline{92.77\%} & \underline{69.87\%} & \underline{72.19\%} & \underline{71.20\%}\\ 
Seed\&Steer\textsubscript{\texttt{gpt-4o}} & & & \textbf{95.80\%} & 69.34\% & \textbf{73.30\%} & \textbf{75.26\%}\\ 
\midrule
\multirow{5}{*}{Seed\&Steer\textsubscript{\texttt{gpt-3.5-turbo}}} 
& Gson & 378 & 83.33\% & 56.61\% & 69.62\% & 66.00\%\\
& Lang & 1728 & 88.60\% & 67.36\% & 65.93\% & 66.46\%\\
& Cli & 177 & 85.88\% & 62.71\% & 63.57\% & 65.69\%\\
& Csv & 137 & 70.00\% & 49.64\% & 65.38\% & 72.19\%\\
& Chart & 5772 & 95.39\% & 72.19\% & 75.55\% & 72.91\%\\ \midrule
\multirow{5}{*}{Seed\&Steer\textsubscript{\texttt{gpt-4o}}} 
& Gson & 378 & 94.71\% & 68.78\% & 76.77\% & 82.06\%\\
& Lang & 1728 & 97.74\% & 77.84\% & 77.60\% & 81.19\%\\
& Cli & 177 & 94.92\% & 66.10\% & 88.10\% & 89.06\%\\
& Csv & 137 & 96.35\% & 79.56\% & 76.75\% & 78.35\\
& Chart & 5772 & 94.65\% & 66.68\% & 71.15\% & 72.70\%\\
\bottomrule
\end{tabular}
\end{table*}

\textbf{\textit{Dataset.}} Defects4J\cite{just2014defects4j} is a widely used benchmark suite that provides real-world software bugs and supporting infrastructure for evaluating software engineering techniques. Following prior work such as ChatUniTest\cite{xie2023chatunitest}, TestART\cite{gu2024improving} and HITS\cite{wang2024hitshighcoveragellmbasedunit}, we select five representative Java projects from Defects4J\cite{just2014defects4j} to evaluate our approach \textit{Seed\&Steer}, as summarized in Table~\ref{datasets}. We utilize the publicly released pickle-format dataset and project environments provided by TestART, which include source code and corresponding pom.xml Maven configuration files. From these environments, we extract all public, non-abstract classes and identify their public methods as focal methods, resulting in a total of \textbf{8,192} focal methods across \textbf{666} classes.

The data we used is not like \textit{A3test}\cite{alagarsamy2024a3test} or \textit{ChatTester}~\cite{yuan2023no} which are in the framework of Defects4J to realize mass runs with different buggy versions of the same project. We chose the same 5 Java projects as them, but fixed the specific versions so that we could compile, execute and evaluate them efficiently. And with a fixed project environment, the EvoSuite-based unit test generation process is easy and repeatable.

\textit{\textbf{Complex Methods.}} To comprehensively evaluate the performance of \textit{Seed\&Steer} on more challenging test generation tasks, we categorize methods under test based on their static complexity and focus on cases with higher structural complexity. as show in Figure~\ref{fig:cc_complexity}. 

\subsection{Baselines}
In order to evaluate the effectiveness of our proposed \textit{Seed\&Steer} method, we use the state-of-the-art unit test generation methods using an LLM as our baseline. Specifically, we choose base model (ChatGPT-3.5, ChatGPT-4.0), ChatUniTest, ChatTester and TestART as our baseline.

\textit{\textbf{ChatUniTest.}} ChatUniTest is a recent LLM-based test generation framework built upon ChatGPT. It adopts a generation-validation-repair paradigm. ChatUniTest is one of the most prominent prompt-based LLM testing frameworks and demonstrates the potential of combining static analysis with in-context learning.

\textit{\textbf{ChatTester.}} ChatTester splits the test generation task into two subtasks firstly understanding the intent of the core method and generating the corresponding unit tests designed to generate tests with higher quality assertions. This is followed by iteratively fixing compilation errors generated during the initial test generation process, maximizing the use of compiler error information to guide LLM.

\textit{\textbf{TestART.}} TestART is an approach to improve LLMbased unit test through the co-evolution of automated generation and repair iteration. TestART leverages the generative capabilities of LLMs by integrating generation-repair co-evolution, testing feedback and prompt injection into the iteration, repairing the bugs contained in the generated test cases and feeding back the coverage information for outputting high-quality test cases.

\subsection{Experiment Setup}
In our evaluation, \textit{Seed\&Steer} performs up to five iterations for each focal method. Within each iteration, it allows at most five rounds of seed repair and five rounds of test case repair. If a generated test case contains compilation or runtime errors that cannot be resolved within the current iteration, the process is terminated early to avoid unnecessary computation. To ensure fair comparison, \textit{Seed\&Steer} and all baseline methods—including TestART —use the \texttt{gpt-3.5-turbo} API, which supports up to 16K tokens of context. For experiments involving \texttt{gpt-3.5-turbo} and \texttt{gpt-4o}, we adopt a unified three-stage procedure: test generation, compilation and execution, and iterative error correction. All hyperparameter settings and prompt formats are kept consistent with those used in \textit{Seed\&Steer}. The evaluation results of \textit{ChatUniTest} are obtained from the official results reported in \textit{TestART}. During test execution, we use Java 8 as the compilation and runtime environment, JUnit 4 as the unit testing framework, and Jacoco\footnote{\url{https://github.com/jacoco/jacoco}} to collect branch and line coverage metrics.

\section{Evaluation And Results}

\begin{figure*}[t]
  \centering
  \includegraphics[width=\textwidth]{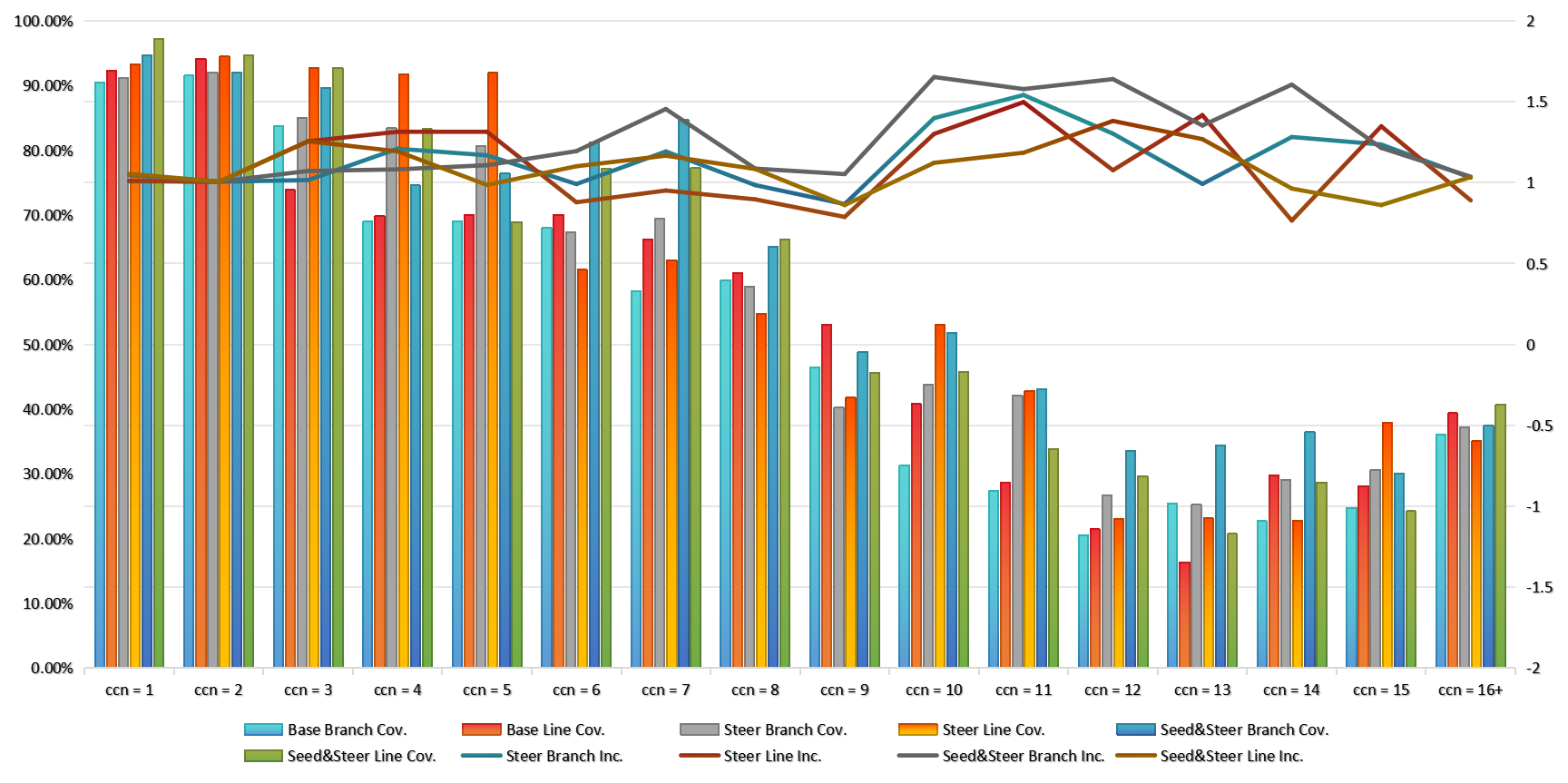}
  \caption{Coverage Improvements from Steer Across Methods of Varying Cyclomatic Complexity}
  \label{fig:steer}
\end{figure*}

\subsection{Research Questions}
We raise the following research questions to evaluate the effectiveness of our method.
\begin{itemize}
    \item \textbf{\textit{RQ1 Effectiveness Comparison.}} How is the performance of Seed\&Steer compared to state-of-the-art (SOTA) approaches?
    \item \textbf{\textit{RQ2 Ablation Comparison.}} Are the components of seed and steer necessary for the success of Seed\&Steer?
    \item \textbf{\textit{RQ3 Compilation Enhancement by Seed.}} Does the seed construction strategy in Seed\&Steer significantly improve the compilation success rate, particularly in addressing cases where an insufficient invocation context causes failures?
    \item \textbf{\textit{RQ4 Coverage Boost by Steer.}} How much does the branch-guided steer strategy contribute to test coverage, and can it consistently improve both branch and line coverage across different levels of method complexity?
\end{itemize}

\begin{table*}[t]
\centering
\caption{Ablation study of \textit{Seed\&Steer} under two LLM backends (\texttt{gpt-3.5-turbo} and \texttt{gpt-4o}).}
\label{tab:ablation-study}
\renewcommand{\arraystretch}{1.2}
\begin{tabular}{lcccccccc}
\toprule
\multirow{2}{*}{\textbf{Variant}} &
\multicolumn{4}{c}{Seed\&Steer\textsubscript{\texttt{gpt-3.5-turbo}}} & 
\multicolumn{4}{c}{Seed\&Steer\textsubscript{\texttt{gpt-4o}}} \\
\cmidrule(lr){2-5} \cmidrule(lr){6-9}
& Compile Pass & Test Pass & Branch Cov. & Line Cov. 
& Compile Pass & Test Pass & Branch Cov. & Line Cov. \\
\midrule
VanillaLLM (ChatTester)     & 85.88\% & 57.18\% & 50.80\% & 48.68\%
                         & 87.21\% & 52.12\% & 52.78\% & 53.52\% \\
+Seed Only               & \textbf{95.54\%} & \textbf{70.47\%} & 59.70\% & 61.09\%
                         & \textbf{98.03\%} & \underline{59.55\%} & 48.82\% & 47.37\% \\
+Steer Only              & 87.95\% & 63.89\% & \underline{61.65\%} & \underline{65.91\%}
                         & 85.68\% & 50.74\% & \underline{68.09\%} & \underline{68.71\%} \\
Seed\&Steer (full)       & \underline{92.77\%} & \underline{69.87\%} & \textbf{72.19\%} & \textbf{71.20\%}
                         & \underline{95.80\%} & \textbf{69.34\%} & \textbf{73.30\%} & \textbf{75.26\%} \\
\bottomrule
\end{tabular}
\end{table*}

\subsection{Answer to RQ1: Effectiveness Comparison}
We evaluated the overall effectiveness of \textit{Seed\&Steer} by comparing it with a set of representative baselines, including \textit{ChatUniTest}, \textit{ChatTester}, \textit{TestART} and raw \textit{ChatGPT-3.5} \& \textit{ChatGPT-4.0}. We report results across multiple dimensions: compilation pass rate, test execution pass rate, branch coverage, and line coverage. 

We compare \textit{Seed\&Steer} with seven baseline approaches in key evaluation metrics. As shown in Table~\ref{tab:Effectiveness comparison}, we first report the average performance in the full data set, followed by a detailed breakdown of the results in five individual Java projects.

Our method demonstrates a significant advantage in terms of compilation pass rate. \textbf{Seed\&Steer's compilation pass rates for calling \texttt{gpt-3.5-turbo} and \texttt{gpt-4o} models are 92.77\% and 95.80\%, respectively.} The direct call to \texttt{gpt-3.5-turbo} (ChatGPT-3.5) compiles with a pass rate of only 74.39\%, while Seed\&Steer(\texttt{gpt-3.5-turbo}) improves the compilation pass rate to 92.77\% . Also calling the \texttt{gpt-3.5-turbo} model, Seed\&Steer(\texttt{gpt-3.5-turbo}) outperforms TestART (+5.72\%), ChatTester (+6.89\%), and ChatUniTest (+16.25\%). Similarly, under \texttt{gpt-4o} model, Seed\&Steer(\texttt{gpt-4o}) achieves 15.30\% and 8.59\% improvements over ChatGPT-4.0 and ChatTester, respectively.

In terms of test execution pass rate, \textit{Seed\&Steer} also achieves competitive results, reaching 69.87\% with \texttt{gpt-3.5-turbo} and 69.34\% with \texttt{gpt-4o}. These values significantly outperform ChatGPT-3.5 (by \~20\%), ChatGPT-4.0 (by ~10\%), and ChatTester (by ~15\%). Although there is still a \~9\% gap compared to TestART (78.55\%), \textit{Seed\&Steer} remains competitive in overall reliability.

For branch and line coverage, \textit{Seed\&Steer} delivers strong performance. When using \texttt{gpt-3.5-turbo}, it achieves 72.19\% branch and 71.20\% line coverage. With \texttt{gpt-4o}, the scores are 73.30\% and 75.26\%, respectively. Compared to the second-best method, TestART (69.40\% and 68.17\%), Seed\&Steer(\texttt{gpt-3.5-turbo}) yields absolute gains of +2.79\% and +3.03\%. The improvements over ChatGPT-3.5, and ChatUniTest are even more pronounced—approaching 30\%, and 25\%, respectively. Notably, Seed\&Steer(\texttt{gpt-4o}) achieves the best overall performance by leveraging the stronger generation capabilities of \texttt{gpt-4o}.
\begin{tcolorbox}[colback=gray!5!white, colframe=black!75!black, 
                  title=Finding 1, fonttitle=\bfseries]

\textit{\textbf{Seed\&Steer}} outperforms existing baselines in terms of compilation pass rate, branch coverage, and line coverage. By leveraging compilable test prefixes and branch-intention-guided assertion generation, it demonstrates superior performance across multiple evaluation metrics.
\end{tcolorbox}

\begin{table*}[t]
\centering
\caption{Compilation Gains With vs. Without Invocation}
\label{tab:seed enhancement}
\renewcommand{\arraystretch}{1.2}
\begin{tabular}{llcccccc}
\toprule
\multirow{2}{*}{\textbf{Model}} &
\multirow{2}{*}{\textbf{Method}} &
\multicolumn{3}{c}{With Invocation} & \multicolumn{3}{c}{Without Invocation}\\
\cmidrule(lr){3-5} \cmidrule(lr){6-8}
& & Invocation num & Pass num & Compile Pass 
& No Invocation num & Pass num & Compile Pass\\
\midrule
\multirow{3}{*}{\textbf{\texttt{gpt-3.5-turbo}}}
& ChatTester\textsubscript{\texttt{gpt-3.5-turbo}} & \multirow{6}{*}{3610} 
                        & 3038 & 84.16\%
            & \multirow{6}{*}{4582} 
                        & 3997 & 87.23\%\\
& SeedOnly                                                      & & +396 & 95.12\% & & +396 & 95.88\%\\
& Seed\&Steer\textsubscript{\texttt{gpt-3.5-turbo}}             & & +284 & 92.02\% & & +281 & 93.37\%\\
\cmidrule(lr){1-2} \cmidrule(lr){4-5} \cmidrule(lr){7-8}
\multirow{3}{*}{\textbf{\texttt{gpt-4o}}}
& ChatTester\textsubscript{\texttt{gpt-4o}}                     & & 3073 & 85.12\% & & 4071 & 88.85\%\\
& SeedOnly                                                      & & +480 & 98.42\% & & +407 & 97.73\%\\
& Seed\&Steer\textsubscript{\texttt{gpt-4o}}                    & & +350 & 94.82\% & & +354 & 96.57\%\\

\bottomrule
\end{tabular}
\end{table*}

\subsection{Answer to RQ2: Ablation Comparison}
To better understand the contribution of each component in \textit{Seed\&Steer}, we conduct an ablation study by selectively disabling the seed construction and steer guidance mechanisms. Specifically, we evaluate the following two variants:
(1) \textit{SeedOnly}, which uses EvoSuite-derived seeds but without branch-guided steer prompts;
(2) \textit{SteerOnly}, which retains the branch-guided prompts but omits seed initialization. In fact the strategy that neither component is used is \textit{ChatTester}. We compare these variants with the full (3) \textit{Seed\&Steer} pipeline across all key metrics to answer RQ2.The results are displayed in Table~\ref{tab:ablation-study}

(1) \textbf{\textit{SeedOnly}}. Adding the Seed component improves compilation pass rates for both models: from 85.88\% to 95.54\% on \texttt{gpt-3.5-turbo}, and from 87.21\% to 98.03\% on \texttt{gpt-4o}. The test execution pass rate also increases notably, from 57.18\% to 70.47\% and from 52.12\% to 59.55\%, respectively. However, coverage rates on \texttt{gpt-4o} decrease slightly with Seed: branch coverage drops from 52.78\% to 48.82\%, and line coverage from 53.52\% to 47.37\%.

(2) \textbf{\textit{SteerOnly}}. The Steer component leads to substantial improvements in coverage. On \texttt{gpt-3.5-turbo}, branch and line coverage increase from 50.80\% and 48.68\% to 61.65\% and 65.91\%, respectively. On \texttt{gpt-4o}, branch coverage improves from 52.78\% to 68.09\%, and line coverage from 53.52\% to 68.71\%. Meanwhile, Steer slightly reduces the compilation pass rate: on \texttt{gpt-3.5-turbo}, it drops from 85.88\% to 87.95\%, and on \texttt{gpt-4o}, from 87.21\% to 85.68\%.

(3) \textbf{\textit{Seed\&Steer}}. Combining both components yields improvements across most metrics. For \texttt{gpt-3.5-turbo}, the compilation pass rate rises to 92.77\%, and branch and line coverage increase to 72.19\% and 71.20\%, outperforming both Seed-only and Steer-only variants. The test pass rate (69.87\%) is slightly below that of Seed alone (70.47\%) but remains comparable. In \texttt{gpt-4o}, compilation success reaches 95.80\% - higher than Steer (85.68\%) but lower than Seed (98.03\%). Coverage improves to 73.30\% (branch) and 75.26\% (line), exceeding both the base and \textit{SeedOnly} methods, and only slightly trailing the \textit{SteerOnly} peak values. 
\begin{tcolorbox}[colback=gray!5!white, colframe=black!75!black, 
                  title=Finding 2, fonttitle=\bfseries]

Ablation studies on the two components of \textit{Seed\&Steer} reveal that each individually contributes to improving LLM performance—compilable seeds enhance compilation pass, while branch-guided assertions improve coverage. However, combining both components can lead to a decrease in compilation pass rate.
\end{tcolorbox}

\subsection{Answer to RQ3: Compilation Enhancement}
To more deeply understand the role of the \textit{seed} component in improving compilation success, we categorize the focal methods into two groups: those containing explicit method invocations and those without. We then evaluate compilation outcomes across three variants—base model, \textit{SeedOnly}, and the full \textit{Seed\&Steer} pipeline on both \texttt{gpt-3.5-turbo} and \texttt{gpt-4o}.

As show in Table~\ref{tab:seed enhancement}, for two base model method, the compilation pass rate is 84.16\% \& 85.12\% for methods with invocations and 87.23\% \& 88.85\% for those without. Interestingly, methods lacking invocations exhibit slightly higher success rates, likely because the model tends to generate minimalistic, syntactically valid test cases in such contexts. 

With the introduction of the Seed component, both categories witness significant improvements. On \texttt{gpt-3.5-turbo}, compilation rates increase to 95.12\% (with-invocation) and 95.88\% (no-invocation), successfully recovering 792 previously failed cases, which accounts for 9.67\% of the dataset. A similar or stronger trend is observed with \texttt{gpt-4o}, where compilation rises from 85.12\% to 98.42\% (with-invocation) and from 88.85\% to 97.73\% (no-invocation), fixing 887 previously failed cases. This validates that Seed component provides richer cases for local methods that contain invocation, and additional cases for local methods that don't contain invocation, and the compilation pass rate is greatly improved. 

Upon full pipeline, we observe a slight decline in compilation success. For \texttt{gpt-3.5-turbo}, the rate drops from 95.54\% (SeedOnly) to 92.77\% (full pipeline), and for \texttt{gpt-4o}, from 98.03\% to 95.80\%. This reduction is expected, as the steer mechanism introduces some prompts that may increase the likelihood of compilation errors. Nevertheless, \textit{Seed\&Steer}'s compilation pass rate is still substantially better than the original model, suggesting that the structural prefixes provided by \textit{Seed} have continued to support overall stability.
\begin{tcolorbox}[colback=gray!5!white, colframe=black!75!black, 
                  title=Finding 3, fonttitle=\bfseries]

Regardless of whether the original class contains an invocation of the focal method, the \textit{Seed} component enriches method invocation examples and helps the LLM compile more test cases. Specifically, it resolves 792 and 887 previously failed cases on the two models, respectively.
\end{tcolorbox}

\begin{table*}[ht]
\centering
\caption{Per-method time cost (in seconds) under different configurations.}
\label{tab:time-cost}
\begin{tabular}{lcccc}
\toprule
\textbf{Project} & \textbf{Seed\&Steer\textsubscript{\texttt{gpt-3.5-turbo}} } & \textbf{ChatTester\textsubscript{\texttt{gpt-3.5-turbo}} } & \textbf{Seed\&Steer\textsubscript{\texttt{gpt-4o}} } & \textbf{ChatTester\textsubscript{\texttt{gpt-4o}}} \\
\midrule
Gson & 27.45 {[}12.18, 37.49{]} & 20.06 {[}13.17, 29.33{]} & 25.92 {[}18.17, 36.16{]} & 19.48 {[}15.17, 29.72{]} \\
Lang & 14.38 {[}9.17, 22.63{]} & 12.62 {[}11.18, 24.47{]} & 12.92 {[}11.18, 15.45{]} & 9.25 {[}8.92, 18.45{]} \\
Cli & 25.21 {[}10.87, 32.16{]} & 17.31 {[}9.87, 29.20{]} & 23.36 {[}15.87, 32.53{]} & 17.05 {[}14.87, 29.71{]} \\
Csv & 38.48 {[}29.25, 47.39{]} & 28.25 {[}23.22, 35.62{]} & 15.90 {[}13.25, 26.82{]} & 10.06 {[}9.34, 18.07{]} \\
JFreeChart & 28.04 {[}23.48, 32.47{]} & 22.56 {[}13.44, 33.31{]} & 26.31 {[}22.12, 38.49{]} & 21.28 {[}15.44, 32.19{]} \\
\midrule
\textbf{Average} & \textbf{25.25 {[}9.17, 47.39{]}} & \textbf{19.43 {[}9.87, 35.62{]}} & \textbf{23.24 {[}11.18, 38.49{]}} & \textbf{19.68 {[}9.34, 32.19{]}} \\
\bottomrule
\end{tabular}
\end{table*}

\begin{table*}[htbp]
\centering
\caption{Performance using classical open source models}
\label{tab:open_source_models_results}
\begin{tabular}{lcccc}
\toprule
\textbf{Method} & \textbf{Compile Passed} & \textbf{Test Passed} & \textbf{Branch Cov.} & \textbf{Line Cov.} \\
\midrule
Seed\&Steer\textsubscript{\texttt{Qwen2.5-Coder-32B-Instruct}} & 88.77\% & 57.51\% & 55.92\% & 50.68\% \\
ChatTester\textsubscript{\texttt{Qwen2.5-Coder-32B-Instruct}}  & 77.16\% & 42.36\% & 44.72\% & 45.78\% \\
Seed\&Steer\textsubscript{\texttt{Qwen2.5-Coder-7B-Instruct}}  & 84.42\% & 48.62\% & 42.91\% & 40.87\% \\
ChatTester\textsubscript{\texttt{Qwen2.5-Coder-7B-Instruct}}   & 75.13\% & 34.53\% & 30.57\% & 27.64\% \\
\bottomrule
\end{tabular}
\end{table*}

\subsection{Answer to RQ4: Coverage Boost}
In our ablation study, we observe that the Steer component significantly improves both branch and line coverage. To further understand \textit{Steer}’s contribution, we grouped focal methods by their cyclomatic complexity number (CCN) using static analysis from \texttt{Lizard}. As shown in Figure~\ref{fig:steer}, \textit{Steer} consistently improves both coverage metrics across all CCN levels under \texttt{gpt-3.5-turbo}. The relative improvement ranges from 1.09× to 1.26×, with more significant gains observed for high-complexity methods (e.g., CCN>9). While methods with CCN between 1 to 4 typically see strong performance from all approaches, more complex cases (e.g., CCN in 11-14) show notable degradation in baseline performance, whereas \textit{Steer} still achieves up to 1.6× coverage improvement.

Compared to the full \textit{Seed\&Steer} pipeline, using \textit{Steer} alone yields more stable and consistent gains across complexity levels. Further analysis of newly covered branches reveals that most of them correspond to deep conditional structures and boundary-specific behaviors, indicating that the intent-driven guidance provided by Steer effectively helps LLMs generate higher-quality assertion statements.
\begin{tcolorbox}[colback=gray!5!white, colframe=black!75!black, 
                  title=Finding 4, fonttitle=\bfseries]

The \textit{Steer} component provides branch-intent context, which enables deeper code coverage even for methods with high cyclomatic complexity, achieving an increase in coverage ranging from 1.09× to 1.26×.
\end{tcolorbox}

\subsection{Threats to Validity}
\textbf{LLM Internal Selection.} Our experiments are conducted under consistent settings, using the same API calls (\texttt{gpt-3.5-turbo} and \texttt{gpt-4o}) and context lengths across all baselines. However, the inherent randomness of LLMs (e.g., temperature or sampling strategies) may influence the generation results. We mitigate this by repeating each test multiple times and reporting averaged outcomes.

\textbf{Datasets Selection.} Our evaluation is limited to five real-world Java projects. Although these projects span a diverse range of complexities and code patterns, they may not fully represent other programming languages or software domains. Future work could explore the generalizability of \textit{Seed\&Steer} to languages such as Python or C++, or to industrial-scale codebases.

\textbf{Metrics Selection.} We adopt four widely used metrics compilation pass rate, test execution pass rate, branch coverage, and line coverage to measure effectiveness. While these metrics are standard in test generation research, they do not directly assess semantic correctness or the adequacy of the generated tests. Future extensions may incorporate human-in-the-loop evaluation or semantic-aware metrics.

\section{Discussion}
\subsection{Execution Overhead.}
\textit{Seed\&Steer} relies on \texttt{EvoSuite} to generate seed tests during initialization. Although this introduces moderate latency when generating the first test for a target class, it remains within a reasonable range compared to other baselines. As shown in Table~\ref{tab:time-cost}, Seed\&Steer exhibits slightly longer average per-method time costs compared to ChatTester (5.8s increase under GPT-3.5 and 3.6s under GPT-4o). However, this overhead is acceptable given the enhanced capability of Seed\&Steer in handling complex initialization contexts and hard-to-cover code paths that are often missed by baseline pipelines. In practice, methods without \texttt{EvoSuite} support often suffer from longer total correction time due to repeated generation of uncompilable code. By providing syntactically and semantically valid seeds upfront, \textit{Seed\&Steer} effectively mitigates this problem.

We do not compare our approach to \texttt{EvoSuite}'s own metrics because its goal is to maximize coverage for the entire class, whereas our approach is oriented towards a specific focal method and \textit{Steer}'s operations cannot be applied to a class.

\subsection{Model Generalization}
Beyond evaluating our approach on proprietary large language models such as OpenAI’s GPT, we also investigate whether the \textit{Seed\&Steer} pipeline generalizes to open-source LLMs. Specifically, we apply our method to Qwen2.5-Coder, a representative open-source code model family, using both the 7B and 32B versions. As shown in Table~\ref{tab:open_source_models_results}, our approach consistently outperforms the \textit{ChatTester}~\cite{yuan2023no} baseline across multiple metrics, including compile success rate and coverage. These results demonstrate that \textit{Seed\&Steer} is not tightly coupled to closed-source models or specific architectures. Instead, it offers a transferable testing generation strategy that is effective even when applied to widely available open-source models.

\subsection{Tool dependency}
A known limitation of \textit{Seed\&Steer} lies in \texttt{EvoSuite}'s strict dependency on Java 8. While Java 8 remains prevalent in many legacy and enterprise systems due to its long-term support, it may not align with modern development environments. However, this constraint does not stem from the design of \textit{Seed\&Steer} itself. Thanks to its modular architecture, \textit{Seed\&Steer} can be easily extended to support newer Java versions by substituting \texttt{EvoSuite} with compatible test generation tools. This flexibility ensures broader applicability across diverse software ecosystems.

And our work demonstrates that it is possible for the output content of traditional tools to be adopted by LLM, and that even without using \texttt{EvoSuite}, the researcher can try to solve problems that we cannot solve on other traditional tools.

\section{Related Work}
\subsection{Automated Unit Test Generation}
% Unit testing is fundamental and critical in the software testing lifecycle, which is significant to ensure the correctness of the software, but the traditional manual design and creation of unit tests is costly and labor-intensive. With the development of technology, automated unit test generation has become a research hotspot. Early automated unit test generation tools mostly use static code analysis methods, such as EvoSuite combines static code analysis and evolutionary search, by accepting Java class or method input, and applying search-based algorithms to generate test suites that satisfy code or branch coverage criteria. However, it suffers from a lack of clarity and readability of the generated tests, and is limited by the Java version.
Automated unit test generation has been a longstanding research topic aimed at alleviating the cost and effort of manual test writing \cite{daka2014survey, klammer2015writing}. Traditional approaches fall into three main categories: search-based, constraint-based, and random-based techniques. Search-based methods apply evolutionary algorithms or heuristics to maximize code coverage, as seen in works like EvoSuite \cite{fraser2010mutation}, TestFul \cite{baresi2010testful}, and others \cite{blasi2022call, delgado2022interevo, harman2009theoretical, derakhshanfar2022basic, harman2001search}. Constraint-based approaches such as DySy \cite{csallner2008dysy}, GRT \cite{ma2015grt}, and Daikon \cite{ernst2007daikon} generate tests by inferring program invariants or solving input constraints. Random-based techniques, like Randoop \cite{pacheco2007feedback} and fuzz testing \cite{zeller2019fuzzing}, generate inputs randomly and rely on statistical exploration. Despite their effectiveness in achieving high coverage, these methods often produce test cases that are hard to interpret, difficult to maintain, and rarely adopted in practice due to their poor readability and limited semantic alignment with developer intent \cite{almasi2017industrial, gargari2021sbst}.

\subsection{Language Model-Based Unit Test Generation}
Recent advances in large language models, especially those based on transformer architectures \cite{vaswani2017attention}, have sparked a surge in research exploring their capabilities for code generation and understanding \cite{feng2024improving, nam2024using, saito2024unsupervised}. There exist transformer-based approaches, such as AthenaTest\cite{tufano2020unit} and A3Test\cite{alagarsamy2024a3test}. ChatUniTest\cite{xie2023chatunitest} adopts a generation-validation-repair paradigm. ChatTESTER\cite{yuan2023no} introduce the intention of method, followed by iterative repair guided by compiler errors. TestART\cite{gu2024improving} further introduces a co-evolutionary loop of test generation and repair, designing some fix templates and integrating feedback signals (e.g., coverage and compilation results) into prompt. HITS\cite{wang2024hitshighcoveragellmbasedunit} semantically splits a method into program slices, guides the LLM to generate unit tests for each slice individually, and finally merges all the tests to improve overall coverage. These incontext-learning pre-processing methods and generic post-processing methods help LLM automated unit test generation. However, these approaches tend to struggle with complex code scenarios, where intricate initialization logic or deeply nested control flows pose challenges to LLMs' understanding and synthesis capabilities.

\section{Conclusion}
We identified two key challenges faced by LLMs in generating unit tests for complex methods through two preliminary experiments. To solve these two challenged, we propose \textit{Seed\&Steer}, a two-phase approach that combines traditional unit testing with the strengths of large language models. In the Seed phase, we leverage \texttt{EvoSuite} to extract method invocation patterns from existing test classes, guiding the LLM to generate valid test prefixes. In the Steer phase, we extract branching conditions to steer the LLM toward generating diverse test intents and assertions covering different execution paths. We conduct experiments on five real-world Java projects, showing that \textit{Seed\&Steer} consistently outperforms existing baselines. In addition, we conducted supplementary experiments to validate the effectiveness and scope of our approach.

We evaluate \textit{Seed\&Steer} on five real-world Java projects against state-of-the-art baselines. Experimental results demonstrate that \textit{Seed\&Steer} improves compilation pass rates by approximately 7\%, successfully compiling 792 and 887 previously failing test cases on two LLMs. It further achieves up to 73\% branch and line coverage across focal methods of varying complexity, with relative coverage improvements ranging from 1.09× to 1.26×. 

We hope that our approach and preliminary study can offer insights and inspiration for future work in LLM-based test generation.

\printbibliography

\end{document}